\documentstyle[prl,aps,epsfig,twocolumn]{revtex}
\begin{document}
\draft
\title{Azimuthal Anisotropy of $\eta$ and $\pi^{0}$ Mesons\\
in Heavy-Ion Collisions at 2 AGeV }

\author{A. Taranenko$^1$, A. Kugler$^1$, R. Pleska\v{c}$^1$, P. Tlust\'y$^1$,
V. Wagner$^1$, H. L\"ohner$^2$, R. W. Ostendorf$^2$, \\
R.H. Siemssen$^2$, P. H. Vogt$^2$, H.W. Wilschut$^2$,
R. Averbeck$^3$, S. Hlav\'{a}\v{c}$^{3,\dagger}$, 
 R. Holzmann$^3$, \\ 
 A.Schubert$^3$, R. S. Simon$^3$, R. Stratmann$^3$, F. Wissmann$^3$, 
Y. Charbonnier$^4$, \\
  G. Mart\'{\i}nez$^{4,5}$, Y. Schutz$^4$, 
J. D\'{\i}az$^5$, A. Mar\'{\i}n$^{5}$, A. D\"{o}ppenschmidt$^6$,
M. Appenheimer$^7$, \\
 V. Hejny$^7$, V. Metag$^{7}$, 
R. Novotny$^7$, H. Str{\"o}her$^7$, 
J. Wei{\ss}$^7$, A.R. Wolf$^7$ and  M. Wolf$^7$}

\address{
$^1$Nuclear Physics Institute, CZ-25068 \v{R}e\v{z}, Czech Republic\\
$^2$Kernfysisch Versneller Instituut, NL-9747 AA Groningen, The Netherlands\\
$^3$Gesellschaft f{\"u}r Schwerionenforschung, D-64291 Darmstadt, Germany\\
$^4$Grand Acc\'{e}l\'{e}rateur National d'Ions Lourds, F-14021 Caen
Cedex, France\\
$^5$Instituto de F\'{\i}sica Corpuscular, 
Centro Mixto Universidad de Valencia--CSIC, E-46100 Burjassot, Spain\\
$^6$Institut f{\"u}r Kernphysik, Universit{\"a}t Frankfurt, D-60486
Frankfurt am Main, Germany\\
$^7$II. Physikalisches Institut, Universit{\"a}t Gie{\ss}en,
D-35392 Gie{\ss}en, Germany}

\date{\today} 

\maketitle

\begin{abstract}
Azimuthal distributions of $\eta$ and $\pi^{0}$ mesons emitted at midrapidity in
collisions of 1.9 AGeV $^{58}$Ni+$^{58}$Ni and  2 AGeV $^{40}$Ca+$^{nat}$Ca are
studied as a function of the number of projectile-like spectator nucleons. The
observed anisotropy corresponds to a negative elliptic flow signal for 
$\eta$ mesons, indicating a
preferred emission perpendicular to the reaction plane. The effect is smallest
in peripheral Ni+Ni collisions. In contrast, for $\pi^{0}$ mesons, elliptic
flow is observed only in peripheral Ni+Ni collisions, changing from positive
to negative sign with increasing pion transverse momentum.

\date{\today} 

\maketitle

\end{abstract}
\narrowtext

Significant compression of nuclear matter achieved during relativistic heavy-ion
collisions manifests 
itself in anisotropies of observed azimuthal distributions of
baryons emerging from the collisions. In-plane emission 
of baryons in the direction of projectile-like (forward hemisphere) or target-like
(backward hemisphere) spectator nucleons is designated as positive directed flow,
while out-of-plane emission of baryons at midrapidity is designated as negative
elliptic flow, see \cite{olit}. \\
Similar to  baryons, negative
elliptic flow was observed for high transverse momentum neutral and 
charged pions  emitted at midrapidity 
in 1 AGeV Au+Au collisions at SIS (GSI)\cite{lars,brill}.
In contrast to positive directed flow
of baryons, directed flow of charged pions was found 
to be negative \cite{kugp94,kint}.
The origin of these anisotropies has not been attributed to the expansion of nuclear matter,
rather to the final 
state interactions of 
 pions with the spectator matter transiently concentrated
in the reaction plane \cite{bao94,bass93}. 
Therefore, an increase in the magnitude of elliptic flow of pions
with increasing size of spectator matter was predicted \cite{bao94}. However, recent 
detailed study of elliptic flow of charged pions  
in Bi+Bi collisions at 0.4-1.0 AGeV reports a very small variation of 
its magnitude  with the change of the number of spectator
nucleons, see \cite{brill97}. 
Hence, the origin of the observed azimuthal anisotropies of 
pions is still unclear. \\
While the main source of pions in relativistic heavy-ion collisions is the
decay of $\Delta$(1232) resonances, $\eta$ mesons are produced mainly by the
decay of N$^*$(1535) resonances. 
The absorption of pions in hot nuclear matter proceeds via the
 $\Delta$(1232) resonance which decays dominantly by pion emission.
However, only 
about 50$\%$  of all N$^*$(1535) resonances created by $\eta$-meson
absorption will reemit $\eta$ mesons. 
Therefore, a comparison of the $\eta$- and $\pi^{0}$  azimuthal anisotropy 
may yield information on the propagation of these  mesons, as well as on the 
dynamics of the parent baryon resonances. \\
Below we present results  of the first experimental study of  
azimuthal distributions of $\eta$ mesons emitted in  
collisions of  1.9 AGeV $^{58}$Ni+$^{58}$Ni and 
2 AGeV  $^{40}$Ca+$^{nat}$Ca nuclei, 
and compare them with azimuthal distributions of $\pi^0$ mesons 
in the same colliding systems. 
The experiments were performed at the Heavy-Ion Synchrotron SIS at GSI 
Darmstadt. \\
 In the first experiment 
a 1.9 AGeV $^{58}$Ni beam with an intensity  
of $6.5\times10^{6}$ particles per spill (spill duration 8~s and repetition 
rate 15~s) was incident on a $^{58}$Ni target (502 mg/cm$^{2}$). In the second
experiment, a $^{nat}$Ca target (320 mg/cm$^{2}$) was bombarded  by a 
$^{40}$Ca beam with kinetic energy 2 AGeV and an intensity of $5\times10^{6}$ particles per
spill (spill duration 10~s and repetition rate 14~s). \\
Photon pairs from the neutral-meson decay were detected in 
the Two-Arm Photon Spectrometer (TAPS) \cite{taps}. This detector system consisted of 
384 BaF$_{2}$ scintillators arranged in 6 blocks of 64 modules with individual 
Charged Particle Veto detectors (CPV) in 
front of each module. The blocks were mounted in two towers 
positioned at 40$^{\circ}$ with respect to the 
beam direction at the distance of 150~cm. Three blocks were positioned in each 
tower at +21$^{\circ}$, 0$^{\circ}$ and -21$^{\circ}$ with respect 
to the horizontal plane. In this setup, only neutral mesons around mid-rapidity $y_{cm}$ were
detected. The geometrical acceptance of TAPS 
for the $\pi^{0}$ and $\eta$ detection was of an order $1\times10^{-3}$. \\
The reaction centrality  was determined by  the hit multiplicity of 
charged particles (M$_{react}$) in a reaction detector. This detector, comprising 
40 small plastic scintillators, was positioned close
to the target and covered the polar angles from 14$^{\circ}$ to 30$^{\circ}$. Most of the 
particles emitted in this angular range are  participant 
nucleons \cite{arka97}. \\
The plastic Forward Wall (FW) of the KaoS collaboration, see \cite{kaos},
comprising 320 plastic scintillators was positioned 520 cm downstream of the target and
covered the polar angles from 
$0.7^{\circ}$ to $10.5^{\circ}$. Particles emitted in this angular range 
are predominantly  projectile-like
spectator nucleons bounced off in the reaction
plane\cite{arka97}. The FW provided the information on emission angle, charge and
time-of-flight of protons and light charged fragments up to Z=8.  \\
The total charge Z$_{FW}$ of particles  detected by the FW 
allowed us to estimate the 
average number of projectile-like spectators $\langle A_{sp}\rangle $ 
for each studied 
bin in multiplicity M$_{react}$,  determined by the reaction detector. 
We used  the relation 
$\langle A_{sp} \rangle=\langle Z_{FW}\rangle A_{proj}/Z_{proj}$,
where $A_{proj}$ and $Z_{proj}$ are mass and charge of the projectile, 
respectively. 
The resulting distributions of the total charge $Z_{FW}$ for studied bins 
in M$_{react}$  are shown in the left column of Fig.1,  
both for the Ni+Ni and Ca+Ca collisions. 
The systematic error of the values $\langle A_{sp} \rangle$ was found to be  less
than 4 units, see \cite{wolf98}.\\
The reaction plane for each event was defined by the incident beam direction and the vector 
\vskip 0.2cm
\begin{equation} 
\vec{Q}=\sum^{n}_{i=1}Z_{i}
\vec{x}_{i}/\mid \vec{x}_{i}\mid ,
\end{equation} 
where the sum runs over all $n$ particles  detected by the FW in the event.
$\vec{x}_{i}$ is the position vector of particle $i$ in the x-y plane in the
FW and $Z_i$ is its charge. The unbiased azimuthal-angle distribution of the 
vector $\vec{Q}$ was found to be flat after introducing a correction, 
which takes into
account the shift of the beam position with respect to the 
geometrical center of the FW. The required  corrections were below 1 cm. \\
Because of finite multiplicity fluctuations, 
 the azimuthal
 angle $\phi$  of the vector $\vec{Q}$ can differ
 from the azimuthal angle of the true reaction plane  $\phi_{true}$ by a deviation
 $\Delta\phi_{pl}$.
 To estimate the width  $\sigma_{pl}$ of 
the distribution $N(\Delta\phi_{pl})$ we randomly divided 
the  hits in each event into two subgroups containing each one half of the number of
particles, see \cite{trmom}. For each subevent one can construct  two independent vectors
$\vec{Q_1}$ and $\vec{Q_2}$, according to Eq.(1) and extract the angle
$\Delta \phi_{12}$=$\phi_{1}-\phi_{2}$ between the two vectors. 
The width $\sigma_{12}$ of 
the resulting distribution $N(\Delta \phi_{12})$   is 
a measure of the precision of the reaction-plane determination. In  
the approximation of a Gaussian distribution of $N(\Delta \phi_{pl})$ 
one has $\sigma_{pl}=\sigma_{12}/2$.
Our analysis was restricted to a sufficient  vector length $\mid\vec{Q}\mid$
 in order to reject the most central events with
no spectator flow. This selection 
 removes  25$\%$ of all registered events. The resolution 
$\sigma_{pl}$ varies between
 $43^{\circ}$ and $55^{\circ}$ depending on the reaction centrality and the colliding system,
 see Table \ref{table1}. These values agree with published data from studies 
of charged-baryon flow in similar colliding systems \cite{trmom,ritman95,prag97}. \\
Photon-particle discrimination has been performed in TAPS as described in \cite{ralf97}. 
For each pair of detected photons in a given event we calculated the
invariant mass $M_{pair}$ and the momenta $\vec{p}_{pair}$
using the following relations:
$M_{pair}^2=2E_{1}E_{2}(1-cos\Theta_{12})$
and
$\vec{p}_{pair}=\vec{p}_{1}+\vec{p}_{2}$ 
\noindent, where $E_{1}$, $\vec{p}_{1}$ and
$E_{2}$, $\vec{p}_{2}$ are the  energies and momenta of the corresponding photons,
$\Theta_{12}$ is the opening angle 
of the photon pair. We analyzed 
only neutral mesons in a narrow rapidity window ($|y-y_{cm}|\le$ 0.1). 
The combinatorial background, due to uncorrelated photon pairs, was deduced by the 
method of event mixing from experimental data, i.e. combining photons from different
events with the same overall event characteristics \cite{arka97,ralf97}. 
The resolution in $M_{pair}$ was 15 MeV/$c^{2}$ 
and 45 MeV/$c^{2}$ (FWHM) for the
$\pi^{0}$ and $\eta$ peaks, respectively \cite{prag97}. \\
We found  that 
only the magnitude of the combinatorial background is dependent
on the azimuthal angle
$\Delta\varphi=\phi_{pair}-\phi$
of meson emission relative to the reaction plane \cite{arka97}.
Therefore, the shape of the combinatorial background deduced by the 
method of event mixing for each bin
in $M_{react}$ and transverse momenta $p_t$ of the photon pair
  was scaled to the experimental data
for each bin in azimuthal angle $\Delta\varphi$ separately and subtracted from the
data. We  present in Fig.1 the resulting 
azimuthal yields of $\eta$ mesons, soft 
$\pi^{0}$ ( 0$\le p_{t} \le 200 $ MeV/c ) and hard $\pi^{0}$ 
( 600$\le p_{t} \le 800 $ MeV/c )
for different bins in M$_{react}$ 
for both  systems studied. \\
We fitted the 
azimuthal yields of $\eta$ and $\pi^0$  mesons both by assuming a constant ( isotropic
distribution ) and 
by the first two terms of a 
Fourier expansion in the azimuthal angle:
\vskip 0.2cm
\begin{equation}
N(\Delta\varphi)=\frac{N_0}{2\pi}\pmatrix{1+2v_1 cos(\Delta\varphi) + 2v_2 cos(2\Delta\varphi)},
\end{equation}
which are used to parametrize the directed ($v_1$) and 
elliptic ($v_2$) flow \cite{vol96}.
The standard $F$-test rejects the fit of azimuthal yields of $\eta$ mesons
 by a constant with a confidence level of
95$\%$. \\ 
The extracted 
values of $v_1$ are 
zero within the error bars, as should be expected  
since 
we study symmetric colliding systems at midrapidity, see  \cite{prag97}. \\
The resulting values of the parameter $v_2$
for $\eta$ and $\pi^0$   mesons
are given in Table \ref{table1}.
 The parameter $v_{2}$ is 
negative for $\eta$ mesons, indicating a preferred emission perpendicular to the
reaction plane (negative elliptic flow). In contrast to previous 
studies of heavy systems at 1 AGeV the elliptic-flow 
signal $v_{2}$ for $\pi^{0}$ mesons is pronounced 
in peripheral ($\langle A_{sp}\rangle=51$) Ni+Ni collisions only. \\
The measured azimuthal distributions are 
affected by the resolution in the
determination of the reaction plane. 
\newpage
\widetext
\begin{figure}
\label{fg:azimuth}
%%%\vspace{8.cm}
\vspace{-0.95cm}
 \begin{center}
    \mbox{
     \epsfxsize=15.213cm
     \epsffile{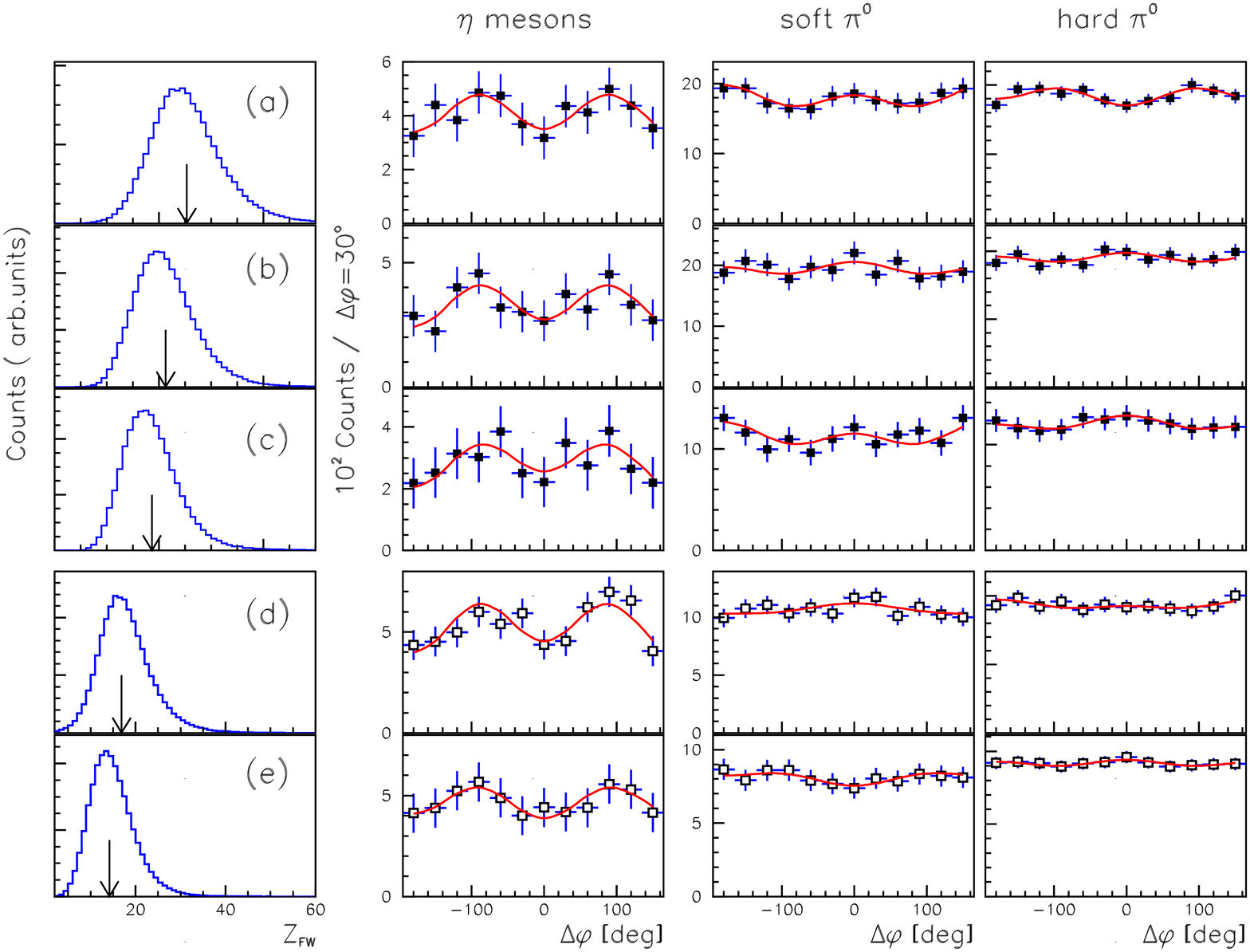}
      }
  \end{center}
  \vspace{-1.0cm}
\begin{center}\parbox{13.6cm}{
\caption{The left part shows the measured distribution of the total charge of 
particles
detected in the FW for different bins in multiplicity M$_{react}$ (see Table \ref{table1}):
(a) - (c) for the experiment  $^{58}$Ni+$^{58}$Ni at 1.9 AGeV and (d) - (e) for the
experiment $^{40}$Ca+$^{nat}$Ca at 2 AGeV. The arrows indicate  mean values
of total charge. 
The right part shows the measured azimuthal angle distributions of $\eta$ mesons, soft 
$\pi^{0}$ (0$\le p_{t} \le 200 $ MeV/c) and  hard $\pi^{0}$ 
(600$\le p_{t} \le 800 $ MeV/c)
 with respect
to the reaction plane which correspond to the same bins in reaction centrality
 (see left part of the picture). The horizontal error bars indicate the bin width.
}
}\end{center}
\vspace{-0.3cm}
\end{figure}
\narrowtext 
Therefore, it is necessary to correct the $v_2$ coefficients
for the fluctuation $\Delta\phi_{pl}=\phi_{true} -\phi$ of the azimuthal angle of the estimated reaction plane with 
respect to the true one.
Averaging over many events,  one obtains the following relation 
between the measured
$v_2$ coefficients and the true $v_{2}^{true}$ coefficients: 
 $v_{2}^{true}$= $v_{2}$/$\langle cos2\Delta\phi_{pl} \rangle$ \cite{trmom,vol96}. 
In our analysis
we used the method described in detail in \cite{olit}, which  allows to
directly extract the values $\langle cos2\Delta\phi_{pl} \rangle$ by means of an
analytical expression from the experimental distribution of  
$N(\Delta \phi_{12})$ (see above).  
The obtained values of $\langle cos2\Delta\phi_{pl} \rangle$ are given in Table \ref{table1}
for each bin in reaction centrality. The systematic error of the 
values $\langle cos2\Delta\phi_{pl} \rangle$ was found to be less than 15$\%$ \cite{prag97}. \\
 The dependence of $v_{2}^{true}$
on the average number of projectile-like spectator nucleons 
$\langle A_{sp} \rangle$ is shown in Fig.2 for
$\eta$ mesons and for two bins in $p_t$ of $\pi^0$ mesons for both
Ni+Ni and Ca+Ca collisions. \\
As it was already  mentioned above, microscopic calculations associate 
elliptic flow of pions with the assumption of their "shadowing" by spectators. 
Guided by this we attempt to describe our data 
by a geometrical model assuming absorption 
of the mesons in the spectator blobs. 
\begin{center}
\end{center}
\vskip 13.61cm
We calculate 
the parameter $v_2^{true}$  as function of the number of spectators A$_{sp}$
from the equation
\begin{equation}
v_2^{true}=0.5(1-R)/(1+R)
\end{equation} 
with $R=exp(L/\lambda)$,
where $\lambda$  is the mean free path for mesons in cold spectator matter and 
$L=2\cdot A_{sp}^{1/3}$~~fm is the mean thickness of spectator matter. 
The mean free path in cold spectator matter for $\eta$ mesons is 
$\lambda_{\eta}\approx$ 1-2 fm for the momentum range
$p_{\eta}\approx$ 50-200 MeV/c  \cite{cass91,mami96},  and for $\pi^0$ mesons 
$\lambda_{\pi^0}\approx$ 1-6
fm \cite{paoli89}. The dash-dotted lines in Fig.2 present the results of
these calculations for two different values of $\lambda$: $\lambda$=2 fm and $\lambda$=6 fm. 
It is obvious, that the simple scenario assuming final-state interactions  
with spectators alone 
can not describe the strong difference between the   $\eta$ 
and $\pi^{0}$ azimuthal anisotropies. The observed dependence of the magnitude
of the azimuthal anisotropy for $\eta$ mesons on the number of spectator
nucleons seems to contradict the model, too. \\
The $\eta$ mesons are solely produced by the decay of the heavy N$^*$(1535) 
resonances which can only be excited in the early stage of the collision. 
Consequently, most of the $\eta$ mesons are emitted after rescattering in the
early phase of the collision, while most of the $\pi^{0}$ mesons are emitted 
after the spectators have passed the collision zone.
This may explain the much stronger azimuthal anisotropy observed for $\eta$ 
mesons. However, for a quantitative explanation microscopic model calculations 
are needed. \\

\begin{figure}
\label{fg:v2true}
%%%\vspace{8.cm}
\vspace{-1.5cm}
 \begin{center}
    \mbox{
     \epsfxsize=8.3cm
     \epsffile{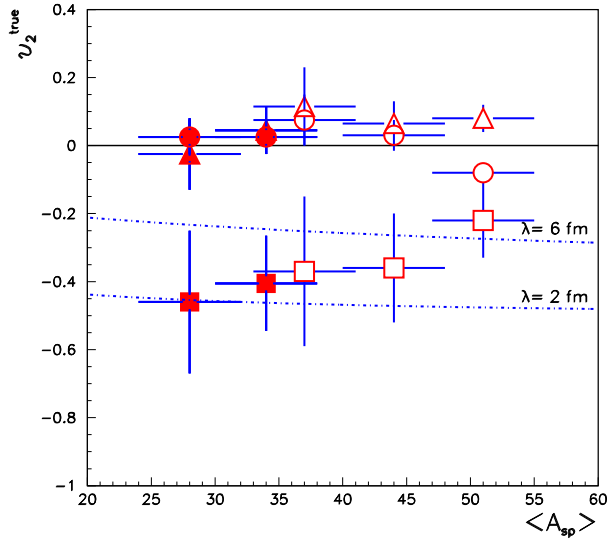}
      }
  \end{center}
  \vspace{-0.55cm}
\begin{center}\parbox{8.4cm}{
\caption{The parameters $v_{2}^{true}$ for $\eta$ mesons (squares),
soft  $\pi^{0}$ ( 0$\le p_{t} \le 200 $ MeV/c ) (triangles) and hard $\pi^{0}$ 
( 600$\le p_{t} \le 800 $ MeV/c ) (circles) as a function
of the number of spectators $\langle A_{sp} \rangle$. Full symbols correspond to Ca+Ca and open
symbols correspond to Ni+Ni collisions. 
The dash-dotted lines represent results of the
 geometrical absorption model (see text).
}
}\end{center}
\vspace{-0.45cm}
\end{figure}
In summary,  we have studied simultaneously azimuthal angular distributions of $\pi^{0}$ 
mesons and of  $\eta$ mesons emitted at midrapidity
in the two colliding systems $^{58}$Ni+$^{58}$Ni at 1.9 AGeV and $^{40}$Ca+$^{nat}$Ca at 2 AGeV.
We observed a strong out-of-plane elliptic flow of 
 $\eta$ mesons. The elliptic flow of $\pi^{0}$ mesons is very weak 
in contrast to  data obtained for heavy colliding systems at 1 AGeV. \\

\widetext
\begin{table}
\caption{Parameters $v_2$ (not corrected for the reaction-plane uncertainty)
 for $\pi^{0}$ and $\eta$ mesons deduced from the experimental azimuthal 
distributions for several intervals (a-e) in M$_{react}$. Corresponding 
mean number of projectile-like spectators $\langle A_{sp} \rangle$, uncertainty of the reaction plane
angle $\sigma_{pl}$, correction of $v_2$ due to this uncertainty are given
(see text). 
For $\pi^0$ mesons the selected intervals in 
transverse momentum p$_t$ are indicated.}
\begin{tabular}{cccccc}
Reaction &\multicolumn{3}{c}{$^{58}$Ni+$^{58}$Ni at 1.9 AGeV} &\multicolumn{2}{c}{$^{40}$Ca+$^{nat}$Ca at 2
AGeV}\\
 M$_{reac}$   & (a) 2 - 6 & (b) 7 - 10 & (c) $\ge$ 11 & (d) 1 - 3 & (e) $\ge$ 4\\ \tableline
$\langle A_{sp} \rangle$  & 51 & 44 & 37 & 34 & 28  \\ 
$\sigma_{pl}$ [$^{\circ}$]   & 43  & 46 & 50 &  52 & 55\\
$\langle cos2\Delta\phi_{pl} \rangle$ & 0.37  & 0.32 & 0.26 &  0.22  &
0.19 \\ \tableline \tableline
~~~~~~~~~~~~~~~~~~~p$_t$ MeV/c &  \multicolumn{5}{c}{}\\ 
$v_{2}(\eta)$~~~~~~~~~~~0 - 600  & -0.08$\pm$0.04 & -0.12$\pm$0.05 & -0.10$\pm$0.06 &  -0.09$\pm$0.03 & -0.09$\pm$0.04 \\
\tableline 
$v_{2}(\pi^{0})$~~~~~~~~~~0 - 200 &  0.033$\pm$0.017 & 0.021$\pm$0.019 & 0.026$\pm$0.031 &  0.012$\pm$0.016&-0.006$\pm$0.021 \\
~~~~~~~~~~~~~~~~~200-400          &  0.005$\pm$0.007 & 0.007$\pm$0.009 & 0.011$\pm$0.016 & -0.011$\pm$0.006&-0.005$\pm$0.011 \\
~~~~~~~~~~~~~~~~~400-600          & -0.004$\pm$0.005 & 0.008$\pm$0.011 &-0.018$\pm$0.016 & 0.006$\pm$0.007 &0.016$\pm$0.007 \\
~~~~~~~~~~~~~~~~~600-800          & -0.031$\pm$0.012 & 0.012$\pm$0.013 & 0.022$\pm$0.021 & 0.005$\pm$0.012 &0.007$\pm$0.012 \\
\end{tabular}
 \label{table1}
 \end{table} 
\narrowtext

This work was supported in part by the Granting Agency of the Czech Republic,
by the Dutch Stichting FOM, the French IN2P3, the German BMBF, 
the Spanish CICYT, by GSI, and the
European Union HCM-network contract ERBCHRXCT94066.

\end{document}